\documentclass[aps,prb,twocolumn,groupedaddress,showpacs]{revtex4}

\usepackage{graphicx}

\begin{document}

\title{Generating spin-entangled electron pairs in 
normal conductors using voltage pulses}

\author{A.V.\ Lebedev$^{\, a}$, G.B.\ Lesovik$^{\, a}$,
and G.\ Blatter$^{\, b}$}

\affiliation{$^{a}$L.D.\ Landau Institute for Theoretical Physics,
RAS, 119334 Moscow, Russia}

\affiliation{$^{b}$Theoretische Physik, ETH-H\"onggerberg, CH-8093
Z\"urich, Switzerland}

\date{\today}

\begin{abstract}
   We suggest an operating scheme for the deliberate generation
   of spin-entangled electron pairs in a normal-metal mesoscopic
   structure with fork geometry. Voltage pulses with associated
   Faraday flux equal to one flux unit $\Phi_0=hc/e$ drive
   individual singlet-pairs of electrons towards the beam splitter.
   The spin-entangled pair is created through a post-selection
   in the two branches of the fork. We analyze the appearance of
   entanglement in a Bell inequality test formulated in terms of
   the number of transmitted electrons with a given spin polarization.
\end{abstract}


\maketitle

\section{Introduction}\label{sec:intro}

Quantum entanglement of electronic degrees of freedom 
in mesoscopic devices has attracted a lot of interest 
recently. Early proposals for structures generating 
streams of entangled particles exploit the interaction 
between electrons as a resource for producing entanglement, 
the pairing interaction in superconductors~\cite{lesovik_01} 
or the repulsive Coulomb interaction in confined 
geometries~\cite{ionicioiu_01,oliver_02}. Recently, 
another class of devices has been suggested which 
avoids direct interparticle interaction; instead, the 
entanglement originates from a proper post-selection of
orbital~\cite{beenakker_03,samuelsson_03,samuelsson_04} or
spin~\cite{fazio_04,lebedev_05,lebedev_04} degrees of 
freedom. The majority of these proposals deals with 
the situation where the entangled particles are emitted 
in a random and uncontrolled fashion, while entanglement
on demand is implicit in the scheme proposed in Ref.\
\onlinecite{ionicioiu_01}.

Current interest concentrates on setups which are 
capable to produce pairs of entangled electrons `on demand'.
Such controlled entanglement is an essential step
towards the realization of quantum computing devices
for which electronic orbital- or spin degrees of freedom
may serve as qubits \cite{loss_98}. In addition,
prospects to convert electronic entanglement into a
photonic one with high efficiency look promising
\cite{loss_04}; this may open new opportunities for the
manipulation of entangled photons with an enhanced
efficiency.

While entanglement on demand is implicit in the work of 
Ionicioiu {\it et al.}, a detailed discussion of the 
controlled production of entanglement in a mesoscopic
device has only been given recently by 
Samuelsson and B\"uttiker \cite{samuelsson_04b}; 
they proposed a scheme for the dynamical generation of 
orbitally entangled electron-hole pairs where a 
time-dependent harmonic electric potential is applied 
between two spatially separated regions of a Mach-Zehnder 
interferometer operating in the Quantum Hall regime.  
This (perturbative) analysis concentrated on the limit 
of a weak pumping potential generating only a small 
fraction of entangled electron-hole pairs per cycle.
Later on, several schemes have been suggested producing
entangled pairs on demand with a high efficiency: in
their setup, Beenakker {\it et al.} \cite{beenakker_05}
make use of a ballistic two-channel conductor driven with
a strong oscillating potential. In their non-perturbative
analysis they demonstrate that this device can pump up
to one (spin- or orbital) entangled Bell-pair per two
cycles. A different proposal based on spin resonance
techniques acting on electrons trapped in a double quantum
dot structure and subsequently released into two quantum
channels has been suggested by  Blaauboer and DiVincenzo
\cite{divincenzo_05}; their detailed analysis of the
manipulation and measurement schemes demonstrates
that the production and measurement of entangled pairs
via an optimal entanglement witness can be performed with
present days experimental technology.
\begin{figure} [h]
   \includegraphics[scale=0.35]{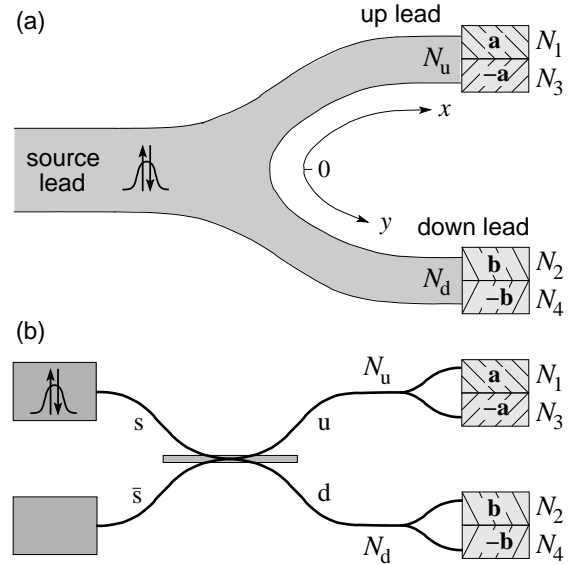}
   \caption[]{Mesoscopic normal-metal structures in a fork geometry
   generating spin-correlated electrons in the two arms of the
   fork. The Bell type setups detect the number of transmitted
   particles $N_i$, $i=1,3$, with spin projected onto the directions
   $\pm{\bf a}$ in the upper arm and correlates them with the number
   $N_j$, $j=2,4$, of particles with a spin projected onto the
   directions $\pm{\bf b}$ in the lower arm. (a) Fork with a simple
   splitter with particles injected along the single source lead.
   (b) Fork in the geometry of a four-terminal beam splitter
   with particles injected along one of the incoming channels
   only. Quenching the transmission $T_\mathrm{ud}$ between
   the upper and lower leads allows to eliminate equilibrium
   fluctuations spoiling the entanglement.}
   \label{fig:fork}
\end{figure}

In the present paper, we discuss an alternative scheme 
generating pulsed spin-entangled electron pairs in a 
normal-metal mesoscopic structure arranged in a fork geometry, 
see Fig.\ \ref{fig:fork}. In this device, spin-entangled 
electron pairs are generated via the injection of spin-singlet
pairs into the source lead from the reservoir \cite{lebedev_05}. 
This entanglement is made accessible by splitting the pair 
into the two leads `u' and `d' and subsequent projection 
(through the Bell measurement) to that part of the wave 
function describing separated electrons travelling in 
different leads \cite{lebedev_05,lebedev_04}.
Rather then quantum pumping with a cyclic potential as
in Refs.\ \onlinecite{samuelsson_04b,beenakker_05}, 
our proposal makes use of definite voltage 
pulses generating spin-entangled electron pairs. A pulsed 
sequence of ballistic electrons is implicitly assumed in the
generation of orbital entanglement by Ionicioiu {\it et al.} 
\cite{ionicioiu_01}, however, no description has been given 
how such (single-electron) pulses are generated in practice. 
Below we discuss a scheme where voltage pulses of specific form 
accumulating one unit of flux $\Phi_0 = -c \int dt \, V(t)$ and 
applied to the source lead `s' generate pairs of spin-entangled
electrons which then are distributed between the two outgoing
leads of the fork, the upper and lower arms denoted as `u' and
`d'. These spin-entangled electron states are subsequently analyzed
in a Bell experiment \cite{bell} involving the measurement of
cross-correlations \cite{nike_02} between the number of electrons
transmitted through the corresponding spin filters in the two arms
of the fork, see Fig.\ \ref{fig:fork}. Using time resolved
correlators, we are in a position to analyze arbitrary forms of
voltage pulses and determine the resulting degree of violation in
the Bell setup. We find that Lorentzian shaped pulses generate
spin-entangled pairs with 50 \% probability, corresponding in
efficiency to the optimal performance of one entangled pair per
two cycles as found by Beenakker {\it et al.} \cite{beenakker_05}.
The reduction in efficiency to 50 \% is due to the competing
processes where the spin-entangled pair generated by the
voltage pulse propagates into only one of the two arms.
In order to make use of this structure as a deterministic
entangler, the Bell measurement setup has to be replaced through a
corresponding projection device (post-) selecting that part of the
wave function with the two electrons distributed between the two
arms; alternatively, this post-selection may be part of the
application device itself, as is the case in the Bell inequality
measurement.

In the following, we first derive (Sec.\ \ref{sec:BI}) an
expression for the Bell inequality involving the particle-number
cross-correlators appropriate for a pulse driven experiment.
We proceed with the calculation of the particle-number
correlators for a single voltage pulse associated with
an arbitrary Faraday flux (Sec.\ \ref{sec:onepulse}). The
results are presented in section \ref{sec:results}: we find
the Bell inequalities violated for single pulses carrying
one Faraday flux, corresponding to one pair of electrons with
opposite spin. Although the Bell inequality appears to be
violated for weak pulses (producing less than one pair) too,
we argue that this violation is unphysical and that its
appearance is due to a misconception in the original derivation
of the Bell inequality arising in the weak pumping limit. We also
generalize the discussion to the situation with more complex
drives (multi-pulse case and alternating pulse sequences)
and demonstrate that our Bell inequalities again are violated
only for single-pair pulses flowing in either direction
through the device. Our analysis of an alternating signal
produces an apparent violation of the Bell inequality,
which, however, again appears to be an artefact resulting
from an improper derivation of the Bell inequality for
the alternating signal. In both cases of failure, weak pulses
and alternating pulse sequences, we encounter backflow
phenomena which spoil the proper derivation of the Bell
inequality for our setup.

\section{Bell Inequality with Number Correlators}\label{sec:BI}

The Bell inequality we are going to use here has been introduced
by Clauser and Horne~\cite{clauser}; it is based on the Lemma
saying that, given a set of real numbers $x$, $\bar x$, $y$, $\bar
y$, $X$, $Y$ with $|x/X|$, $|\bar x/X|$, $|y/Y|$, and $|\bar y/Y|$
restricted to the interval $[0,1]$, the inequality $|xy - x\bar y
+\bar x y + \bar x \bar y|\leq 2|XY|$ holds true. We define the
operator of electric charge $\hat N_i(t_\mathrm{ac})$ transmitted
through the $i$-th spin detector during the time interval
$[0,t_\mathrm{ac}]$, where $t_\mathrm{ac}>0$ is the accumulation
time. The charge operator $\hat N_i(t_\mathrm{ac})$ can be
expressed via the electric current $\hat I_i(t)$ flowing through
the $i$-th detector, $\hat N_i(t_\mathrm{ac})= \int_0^{t_\mathrm{ac}}
dt^\prime\, \hat I_i(t^\prime)$. In the Bell test experiment, see
Fig.~\ref{fig:fork}, one measures the number of transmitted
electrons with a given spin polarization, $N_i$, $i=1,\dots,4$,
and defines the quantities $x=N_1-N_3$, $y=N_2-N_4$, $X=N_1+N_3$,
and $Y=N_2+N_4$ for fixed orientations ${\bf a}$ and ${\bf b}$ of
the polarizers (and similar for $\bar x$ and $\bar y$ for the
orientations $\bar {\bf a}$ and $\bar{\bf b}$), see
Ref.~\onlinecite{nike_02}. Our Bell setup measures the correlations
\begin{eqnarray}
      {\cal K}_{ij}({\bf a},{\bf b}) &=&
      \langle \hat N_i(t_\mathrm{ac}) \hat N_j(t_\mathrm{ac}) \rangle
      \nonumber\\
      &=& \int\limits_0^{t_\mathrm{ac}} dt_1 dt_2 \,
      \langle \hat I_i(t_1) \hat I_j(t_2) \rangle
      \label{ncor}
\end{eqnarray}
between the number of transmitted electrons $N_{i}$, $i=1,3$, in
the lead `u' with spin polarization along $\pm{\bf a}$ and
their partners $N_j$, $j=2,4$, in lead `d' with spin
polarization along $\pm{\bf b}$. Using the above definitions
for $x$, $y$, $X$, and $Y$, we obtain the normalized
particle-number difference correlator,
\begin{eqnarray}
      E({\bf a},{\bf b}) &=&
      \frac{\langle [\hat N_1 -\hat N_3][\hat N_2 -\hat N_4]\rangle}
      {\langle [\hat N_1 +\hat N_3][\hat N_2 +\hat N_4]\rangle}
      \nonumber\\
      &=& \frac{{\cal K}_{12}-{\cal K}_{14}-{\cal K}_{32}+{\cal K}_{34}}
      {{\cal K}_{12}+{\cal K}_{14}+{\cal K}_{32}+{\cal K}_{34}},
\end{eqnarray}
and evaluating the correlators for the four different combinations
of directions ${\bf a},~\bar{\bf a}$ and ${\bf b},~\bar{\bf b}$,
we arrive at the Bell inequality
\begin{equation}
      E_{\scriptscriptstyle\rm BI} =
      | E({\bf a},{\bf b}) - E({\bf a},\bar {\bf b})
      + E(\bar {\bf a},{\bf b}) + E(\bar {\bf a},\bar {\bf b}) |
      \leq 2.
      \label{BI1}
\end{equation}

We proceed further by separating the current
correlators in Eq.~(\ref{ncor}) into irreducible parts
$C_{ij}({\bf a},{\bf b}; t_1,t_2)=\langle \delta \hat I_i(t_1)
\delta \hat I_j(t_2) \rangle$ with $\delta\hat I_i(t)= \hat
I_i(t)- \langle \hat I_i(t)\rangle$ and products of average
currents and rewrite $E({\bf a},{\bf b})$ in the form
\begin{equation}
      E({\bf a},{\bf b}) =
      \frac{K_{12} - K_{14} - K_{32} + K_{34} + \Lambda_-}
      {K_{12} + K_{14} + K_{32} + K_{34}+\Lambda_+},
\end{equation}
where we have defined $\Lambda_\pm = [\langle \hat{N_1} \rangle
\pm \langle \hat{N_3} \rangle][\langle \hat{N_2}\rangle \pm
\langle \hat{N_4} \rangle]$ with the irreducible particle
number correlator
\begin{eqnarray}
   \label{K}
   K_{ij}(t_\mathrm{ac})
   &=& \langle \delta\hat N_i(t_\mathrm{ac})
   \delta\hat N_j(t_\mathrm{ac})\rangle \\
   &=& \int_0^{t_\mathrm{ac}} dt_1 dt_2 \,
   C_{ij}({\bf a},{\bf b};t_1,t_2).
   \nonumber
\end{eqnarray}
The average currents are related via $\langle \hat
I_1(t) \rangle=\langle\hat I_3(t)\rangle=\langle\hat I_\mathrm{u}
(t)\rangle/2$ and $\langle \hat I_2(t)\rangle =\langle \hat
I_4(t)\rangle =$ $\langle \hat I_\mathrm{d}(t)\rangle/2$ and thus
$\Lambda_-=0$, $\Lambda_+=\langle \hat N_\mathrm{u}\rangle \langle
\hat N_\mathrm{d}\rangle$. The irreducible current-current
correlator factorizes into a product of spin and orbital parts,
$C_{ij}({\bf a},{\bf b};t_1,t_2) =|\langle {\bf a}_i|{\bf b}_j
\rangle|^2 C_\mathrm{ud}(t_1,t_2)$ with ${\bf a}_{1,3}= \pm{\bf
a}$ and ${\bf b}_{2,4}=\pm{\bf b}$. The spin projections involve
the angle $\theta_{{\bf a}{\bf b}}$ between the directions $\bf a$
and $\bf b$ of the polarizers, $\langle \pm {\bf a}|\pm {\bf
b}\rangle = \cos^2 (\theta_{{\bf a}{\bf b}}/2)$ and $\langle \pm
{\bf a}|\mp {\bf b}\rangle = \sin^2 (\theta_{{\bf a}{\bf b}}/2)$,
and the Bell inequality assumes the form
\begin{equation}
      \left|\frac{K_\mathrm{ud}
      [\cos\theta_{{\bf a}{\bf b}}-\cos\theta_{{\bf a}\bar{\bf b}}
      +\cos\theta_{\bar{\bf a}{\bf b}}+\cos\theta_{\bar{\bf a}\bar{\bf b}}]}
      {2K_\mathrm{ud}+\langle \hat N_\mathrm{u}\rangle \langle
      \hat N_\mathrm{d}\rangle}
      \right|\leq1,
      \label{BI2}
\end{equation}
where $K_\mathrm{ud}(t_\mathrm{ac})= \int_0^{t_\mathrm{ac}} dt_1
dt_2\, C_\mathrm{ud}(t_1,t_2)$ is the (irreducible) number
cross-correlator between the upper and lower leads of the fork.
The maximal violation of the Bell inequality is attained for the
standard orientations of the detector polarizations $\theta_{{\bf
a}{\bf b}}=\theta_{\bar{\bf a}{\bf b}} =\theta_{\bar{\bf
a}\bar{\bf b}}=\pi/4$, $\theta_{{\bf a}\bar{\bf b}} =3\pi/4$; the
Bell inequality (\ref{BI2}) then reduces to
\begin{equation}
      E_{\scriptscriptstyle\rm BI} = \left|
      \frac{2K_\mathrm{ud}}{2K_\mathrm{ud}+\langle \hat N_\mathrm{u}
      \rangle \langle \hat N_\mathrm{d}\rangle}\right| \leq
      \frac1{\sqrt2}.
      \label{BI3}
\end{equation}

\section{Number Correlators for a Single Pulse}\label{sec:onepulse}

The orbital part $C_\mathrm{ud}(t_1,t_2)$ of the current
cross-corre\-lator between the upper and lower leads can be
calculated within the standard scattering theory of noise
\cite{lesovik_89,buttiker_90,lesovik_99,blanter}. We assume
that the time dependent voltage drop $V(t)$ at the splitter can
be treated adiabatically (i.e., the voltage changes slowly
during the electron scattering time). The electrons incident
from the source lead `s' and scattered to the `up' or `down'
lead then acquire an additional time dependent phase $\phi(t)
=\int_{-\infty}^t dt^\prime\, eV(t^\prime)/\hbar$. The
scattering states (for one spin component) describing the
electrons in the upper and lower leads, $\hat \Psi_\mathrm{u}
(x,t)$ and $\hat \Psi_\mathrm{d}(x,t)$, take the form
\begin{eqnarray}
      &&\hat \Psi_\mathrm{u} =
      \int \frac{d\epsilon}{\sqrt{h v_\epsilon}}\Bigl[
      \bigr( t_\mathrm{su} e^{i\phi(t-x/v_\epsilon)}
      \hat c_\epsilon + r_\mathrm{u} \hat a_\epsilon +
      t_\mathrm{du}\hat b_\epsilon \bigr) e^{ikx}
      \nonumber\\
      && \qquad\qquad
      +\,\hat a_\epsilon e^{-ikx} \Bigr] e^{-i\epsilon t/\hbar},
      \label{scu}
      \\
      &&\hat \Psi_\mathrm{d} =
      \int \frac{d\epsilon}{\sqrt{h v_\epsilon}}\Bigl[
      \bigr( t_\mathrm{sd} e^{i\phi(t-x/v_\epsilon)}
      \hat c_\epsilon + r_\mathrm{d} \hat b_\epsilon +
      t_\mathrm{ud}\hat a_\epsilon \bigr) e^{ikx}
      \nonumber\\
      && \qquad\qquad
      +\,\hat b_\epsilon e^{-ikx} \Bigr] e^{-i\epsilon t/\hbar},
      \label{scd}
\end{eqnarray}
where $v_\epsilon = \sqrt{2m\epsilon}$; $\hat a_\epsilon$, $\hat
b_\epsilon$, and $\hat c_\epsilon$ denote the annihilation
operators for spinless electrons at energy $\epsilon$ in leads
`u', `d', and `s'; the scattering amplitudes $t_\mathrm{su}$
($t_\mathrm{du}$) and $t_\mathrm{sd}$ ($t_\mathrm{ud}$) describe
particle transmission from the source (down) lead into the upper
lead and from the source (up) lead into the lower (`d') lead;
$r_\mathrm{u}$, $r_\mathrm{d}$ denote the reflection amplitudes
into the leads `u' and `d'. Such adiabatically deformed scattering
states (\ref{scu}) and (\ref{scd}) have first been used
in the calculation of the spectral noise power in an $ac$-driven
system \cite{ll_94}; the validity of this approach has been
confirmed in several experiments \cite{ksp_00}.

We substitute these expressions into the current operator
$\hat I_\mathrm{u(d)}(x,t)$ and drop all terms small in
the parameter $|\epsilon-\epsilon^\prime|/
\epsilon_{\scriptscriptstyle\rm F}$ (we assume a linear
dispersion). The irreducible current cross-correlator
$C_\mathrm{ud}(t_1,t_2) = \langle \delta \hat I_\mathrm{u}
(x,t_1) \delta \hat I_\mathrm{d} (y,t_2) \rangle$ measured
at the positions $x$ and $y$ in the leads `u' and `d' can
be splitted into two terms, one due to equilibrium
fluctuations, $C_\mathrm{ud}^\mathrm{eq} (t_1-t_2)= \int
(d\omega/2\pi)\,S^\mathrm{eq}(\omega)e^{i\omega(t_1-t_2)}$ with
\begin{equation}
      S^\mathrm{eq}(\omega) = -\frac{2e^2}{h}\, T_\mathrm{ud}
      \cos(\omega\tau^+)\,\frac{\hbar\omega}{1-e^{\hbar\omega/\theta}},
\end{equation}
and a second term describing the excess correlations at finite
voltage,
\begin{equation}
      C_\mathrm{ud}^\mathrm{ex}(t_1,t_2) = -\frac{4e^2}{h^2}
      T_\mathrm{u} T_\mathrm{d}
      \sin^2 \frac{\phi(\xi_1)\!-\!\phi(\xi_2)}2\,
      \alpha(\tau\!-\!\tau^-,\theta),
      \label{cor_eq_ex}
\end{equation}
with $\alpha(\tau,\theta)= \pi^2\theta^2 / \sinh^2 [ \pi\theta\tau
/ \hbar]$ ($\theta$ is the temperature of electronic reservoirs),
$\tau=t_1-t_2$, $\tau^\pm =(x\pm y)/v_{\scriptscriptstyle\rm F}$,
$\xi_1 = t_1-x/v_{\scriptscriptstyle\rm F}$, and
$\xi_2=t_2-y/v_{\scriptscriptstyle\rm F}$. The coefficients
$T_\mathrm{u} = |t_\mathrm{su}|^2$, $T_\mathrm{d} =
|t_\mathrm{sd}|^2$, and $T_\mathrm{ud} = |t_\mathrm{ud}|^2 =
|t_\mathrm{du}|^2$ denote the transmission probabilities from the
source to the `up', `down' leads, and from the `down' to the `up'
lead.

The equilibrium part of the current cross-correlator
$C_\mathrm{ud}^\mathrm{eq}(t_1-t_2)$ describes the correlations
of the electrons in the Fermi sea propagating ballistically
from lead `u' to lead `d' (or vice versa) with the retardation
$\tau^+=(x_1+x_2)/v_{\scriptscriptstyle\rm F}$. The corresponding
equilibrium part of the particle-number cross-correlator,
$K_\mathrm{ud}^\mathrm{eq} = \int_0^{t_\mathrm{ac}} dt_1 dt_2
\, C_\mathrm{ud}^\mathrm{eq}(t_1-t_2)$ then takes the form
\begin{equation}
      K_\mathrm{ud}^\mathrm{eq} \approx \frac{e^2}{\pi^2}
      T_\mathrm{ud} \ln \frac{t_\mathrm{ac}}{\tau},\quad
      \tau = \mbox{max}\{\hbar/\epsilon_{\scriptscriptstyle\rm
      F}, \tau^+\}, \label{ncceq}
\end{equation}
where we have assumed the zero temperature limit and an accumulation
time $t_\mathrm{ac}\gg\tau$. The logarithmic divergence in
$t_\mathrm{ac}$ reduces the violation of the Bell inequality
Eq.~(\ref{BI3}) at large accumulation times and one has
to suppress the equilibrium correlations between the upper
and the lower leads in the setup. This can be achieved via
a reduction in the transmission probability $T_\mathrm{ud}$,
however, in the fork geometry of Fig.~\ref{fig:fork}(a) the
probability $T_\mathrm{ud}$ cannot be made to vanish.
Alternatively, one may chose a setup with a reflectionless
four-terminal beam splitter as sketched in Fig.~\ref{fig:fork}(b)
with no exchange amplitude between the upper and lower outgoing
leads; using such a fork geometry, the equilibrium fluctuations
$K_\mathrm{ud}^\mathrm{eq}$ can be made to vanish 
\cite{note}.

Next, we concentrate on the excess part $K_\mathrm{ud}^\mathrm{ex}$
of the particle-number cross-correlator $\langle \hat N_\mathrm{u}
(t_\mathrm{ac}) \hat N_\mathrm{d}(t_\mathrm{ac}) \rangle $.
Note that the excess fluctuations are the same for both setups
Fig.~\ref{fig:fork}(a) and (b) and we can carry out all the
calculations for the fork geometry. We consider a sharp voltage
pulse applied at time $t_0$, $0<t_0<t_\mathrm{ac}$, with short
duration $\delta t$. The total accumulated phase $\phi(t)$ then
exhibits a step-like time dependence with the step height
$\Delta\phi= \phi(t_0 +\delta t/2)-\phi(t_0-\delta t/2)=
-2\pi \Phi/\Phi_0$, where we have introduced the Faraday
flux $\Phi=-c\int V(t) dt$ and $\Phi_0 = hc/e$ is the flux
quantum. The excess part of the particle-number cross-correlator
$K_\mathrm{ud}$ then takes the form (we consider again the zero
temperature limit)
\begin{equation}
      K_\mathrm{ud}^\mathrm{ex} = -
      \frac{e^2}{\pi^2}T_\mathrm{u} T_\mathrm{d}\!\!
      \int\limits_0^{t_\mathrm{ac}} \!\! dt_1 dt_2\,
      \frac{\sin^2[(\phi(t_1)-\phi(t_2))/2]}{(t_1-t_2)^2}.
      \label{ncord}
\end{equation}
For a sharp pulse with $\delta t \ll t_0, t_\mathrm{ac}$ we can
identify two distinct contributions arising from the integration
domains $|t_1-t_2|\ll \delta t$ and $|t_1-t_2|\gg \delta t$,
cf.\ Refs.\ \onlinecite{ll_93} and \onlinecite{lll_96};
we denote them with $K^<$ and $K^>$. Introducing the average
and relative time coordinates $t=(t_1+t_2)/2$ and $\tau=t_1-t_2$
and expanding the phase difference $\phi(t_1)-\phi(t_2)=
\phi(t+\tau/2)-\phi(t-\tau/2) \approx \dot \phi(t)\tau$, the
first contribution $K^<$ reads
\begin{eqnarray}
      K^< &=& -\frac{e^2}{\pi^2} T_\mathrm{u} T_\mathrm{d}
      \int\limits_0^{t_\mathrm{ac}} dt
      \int d\tau \, \frac{\sin^2[\dot \phi(t)\tau/2]}{\tau^2}
      \nonumber\\
      &=& -\frac{e^2}{2\pi} T_\mathrm{u} T_\mathrm{d}
      \int\limits_0^{t_\mathrm{ac}} dt\, |\dot \phi(t)|.
\end{eqnarray}
Assuming that the phase $\phi(t)$ is a monotonic function of $t$
(guaranteeing a unique sign for $\dot \phi(t)$) the last equation
can be rewritten in terms of the Faraday flux $\Phi$,
\begin{equation}
      K^< = -e^2
      T_\mathrm{u} T_\mathrm{d}\,\frac{|\Phi|}{\Phi_0};
\end{equation}
this contribution to the particle-number cross-correlator
$K_\mathrm{ud}^\mathrm{ex}$ describes the correlations
arising from the $n=|\Phi|/\Phi_0$ additional particles
pushed through the fork by the voltage pulse $V(t)$,
see Eq.\ (\ref{N_Phi}) below.

The second contribution $K^>$ to $K_\mathrm{ud}^\mathrm{ex}$
originates from the time domains $0<t_{1(2)}<t_0-\delta t/2$
and $t_0+\delta t/2<t_{2(1)}<t_\mathrm{ac}$, where $|\phi(t_1)
-\phi(t_2)|=2\pi \Phi/\Phi_0$, hence
\begin{equation}
      K^> \approx
      -\frac{2 e^2}{\pi^2} T_\mathrm{u} T_\mathrm{d} \sin^2
      \frac{\pi\Phi}{\Phi_0}\ln \frac{t_\mathrm{m}}{\delta t};
      \label{response}
\end{equation}
here, we have kept the most divergent term in the measurement
time $t_\mathrm{m}=t_\mathrm{ac}-t_0$, the time during which
the pulse manifests itself in the detector. The above
expression describes the response of the electron
gas to the sudden perturbation $V(t)$; the logarithmic
divergence in the measurement time $t_\mathrm{m}$ can be
interpreted \cite{ll_93} along the lines of the
orthogonality catastrophe \cite{anderson}, with the
isolated perturbation in space, the impurity, replaced
by the sudden perturbation in time. The periodicity of
the response in the Faraday flux $\Phi$ is due to the
discrete nature of electron transport as expressed
through the binomial character of the distribution
function of transmitted particles \cite{ll_93,lll_96}.
Remarkably, the above logarithmically divergent
contribution to $K_\mathrm{ud}^\mathrm{ex}$ vanishes
for voltage pulses carrying an integer number of electrons
$n=|\Phi|/\Phi_0$, see (\ref{N_Phi}) below. This follows
quite naturally from the invariance of the scattering
amplitudes $t_\mathrm{su}$ and $t_\mathrm{sd}$ in
Eqs.~(\ref{scu}) and (\ref{scd}) under the (adiabatic)
voltage pulses carrying integer flux $\pm n \Phi$,
$t_\mathrm{sx} \rightarrow t_\mathrm{sx} e^{\pm 2\pi n}$
with  $\mathrm{x}=\mathrm{u,d}$; transmitting an integer
number of particles at Faraday fluxes $\Phi = n \Phi_0$
avoids the system shakeup and the associated logarithmic
divergence.

We proceed with the determination of the average number of
transmitted (spinless) particles $\langle \hat N_\mathrm{u(d)}
(t_\mathrm{ac}) \rangle = \int_0^{t_\mathrm{ac}} dt\,
\langle \hat I_\mathrm{u(d)}(x,t) \rangle$. Within the
scattering matrix approach the average currents in the
upper and lower leads are given by the expression
\begin{eqnarray}
      \langle \hat I_\mathrm{u(d)}(x,t) \rangle &=& \frac{e}{h}
      T_\mathrm{u(d)}\, eV(t-x/v_{\scriptscriptstyle\rm F})
      \nonumber\\
      &=&\frac{e}{2\pi} T_\mathrm{u(d)}
      \dot \phi(t-x/v_{\scriptscriptstyle\rm F}).
      \label{Iud}
\end{eqnarray}
The time integration provides the average number of
transmitted particles
\begin{equation}
      \langle \hat N_\mathrm{u(d)}(t_\mathrm{ac})\rangle
      = e T_\mathrm{u(d)}\,\frac{\Phi}{\Phi_0}.
      \label{N_Phi}
\end{equation}
With $T_\mathrm{u}+T_\mathrm{d}=1$, the result (\ref{N_Phi}) tells
that a voltage pulse corresponding to $n=|\Phi|/\Phi_0$ flux units
pushes $n$ spinless electrons through the fork, in forward direction
from the source lead `s' to the prongs `u' and `d' if $\Phi>0$ and
in the backward direction for $\Phi<0$.

\section{Results}\label{sec:results}

Substituting the above expressions for the particle-number
cross-correlators and for the average number of transmitted particles
into (\ref{BI3}) we arrive at the following general result for
the Bell inequality
\begin{equation}
      E_{\scriptscriptstyle\rm BI} =
      \left|
      \frac{n+(2/\pi^2)\sin^2(\pi n) \ln(t_\mathrm{m}/\delta t)}
      {2n^2-n-(2/\pi^2)\sin^2(\pi n) \ln(t_\mathrm{m}/\delta t)}
      \right|.
      \label{BI4}
\end{equation}

\subsection{Pulse with integer flux}\label{sec:if}

For a voltage pulse with integer $n$ the above expression simplifies
dramatically as all logarithmic terms vanish, leaving us with
the Bell inequality
\begin{equation}
      E_{\scriptscriptstyle\rm BI} =
      \left|
      \frac{1}{2n-1}
      \right|\leq\frac1{\sqrt{2}},
      \label{BI5}
\end{equation}
which we find maximally violated for $n=1$ and never violated
for larger integers $n>1$ --- any additional particle
accumulated in the detector spoils the violation of the Bell
inequality. Furthermore, this violation is independent of
the transparencies $T_\mathrm{u}$, $T_\mathrm{d}$ and hence
universal; moreover, the Bell inequality~(\ref{BI5}) does
not depend on the particular form or duration of the applied
voltage pulse but involves only the number of electrons $n$
carried by the voltage pulse.

A voltage pulse with $n=1$ pushes two electrons with opposite spin
polarization towards the beam splitter. Such a pair appears in a
singlet state \cite{lebedev_05}
and can be described by the wave
function $\Psi_\mathrm{in}^{\scriptscriptstyle 12} =
\phi_\mathrm{s}^{\scriptscriptstyle 1}
\phi_\mathrm{s}^{\scriptscriptstyle 2}
\chi_\mathrm{sg}^{\scriptscriptstyle 12}$ with the spin-singlet
state $\chi_\mathrm{sg}^{\scriptscriptstyle 12} =
[\chi_{\uparrow}^{\scriptscriptstyle 1}
\chi_{\downarrow}^{\scriptscriptstyle 2} -
\chi_{\downarrow}^{\scriptscriptstyle 1}
\chi_{\uparrow}^{\scriptscriptstyle 2}]/\sqrt{2}$;
$\phi_\mathrm{s}$ is the orbital part of the wave function
describing a particle in the source lead `s' and the upper indices
1 and 2 denote the particle number. This local spin-singlet pair
is scattered at the splitter and the wave function
$\Psi_\mathrm{in}^{\scriptscriptstyle 12}$ transforms to
$\Psi_\mathrm{scat}^{\scriptscriptstyle 12} = t_\mathrm{su}^2
\phi_\mathrm{u}^{\scriptscriptstyle 1}
\phi_\mathrm{u}^{\scriptscriptstyle 2}
\chi_\mathrm{sg}^{\scriptscriptstyle 12}+ t_\mathrm{sd}^2
\phi_\mathrm{d}^{\scriptscriptstyle 1}
\phi_\mathrm{d}^{\scriptscriptstyle 2}
\chi_\mathrm{sg}^{\scriptscriptstyle 12}+ t_\mathrm{su}
t_\mathrm{sd} [\phi_\mathrm{u}^{\scriptscriptstyle 1}
\phi_\mathrm{d}^{\scriptscriptstyle 2} +
\phi_\mathrm{d}^{\scriptscriptstyle 1}
\phi_\mathrm{u}^{\scriptscriptstyle
2}]\chi_\mathrm{sg}^{\scriptscriptstyle 12}$, where the last term
describes two particles in a singlet state shared between the
upper and lower leads of the fork. The Bell inequality test is
only sensitive to pairs of particles propagating in different
arms, implying a projection of the scattered wave function
$\Psi_\mathrm{scat}^{\scriptscriptstyle 12}$ onto the spin-entangled
component. Thus the origin of the entanglement is found in the
post-selection during the cross-correlation measurement effectuated
in the Bell inequality test~\cite{samuelsson_04,lebedev_04}.
From an experimental point of view it may be difficult to produce
voltage pulses driving exactly one (spinless) particle $n=1$.
However, as follows from the full expression Eq.\ (\ref{BI4}),
for a sufficiently small deviation $|\delta n| =|n-1| \ll 1$ the
logarithmic terms are small in the parameter $(\delta n)^2$ and
thus can be neglected, provided the measurement time $t_m$
satisfies the condition $(\delta n)^2\ln (t_\mathrm{m}/\delta
t)\ll 1$.

\subsection{Weak pumping regime}\label{sec:wp}

The weak pumping regime with $n < 1$, corresponding to small
voltage pulses carrying less then one electron per spin
channel, deserves special attention. Inspection of (\ref{BI4})
shows that the Bell inequality can be formally violated in this
regime. We believe that this violation of the Bell inequality
has no physical meaning. Below, we will show that for non-integer
$n$ many-particle effects generate backflow of particles in our
setup. We argue that this backflow leads to an inconsistency
in the derivation of the Bell inequality itself, as the
assumption $0\leq |x/X|,|\bar x/X|, |y/Y|,|\bar y/Y| \leq 1$
may no longer hold true for $n <1$.

In order to understand this weak pumping regime better, we analyze
the sign of the full current cross-correlator ${\cal C}_\mathrm{ud}
(t_1,t_2)=\langle \hat I_\mathrm{u}(t_1) \hat I_\mathrm{d}(t_2)
\rangle$ (for one spin component); expressing this quantity
through the phase $\phi(t)$ we find the form
\begin{eqnarray}
      &&{\cal C}_\mathrm{ud}(t_1,t_2) =
      \frac{e^2}{(2\pi)^2} T_\mathrm{u} T_\mathrm{d}
      \nonumber\\
      &&\times
      \left[
      \dot \phi(t_1) \dot \phi(t_2) -4
      \frac{\sin^2[(\phi(t_1)-\phi(t_2)/2]}{(t_1-t_2)^2}
      \right].
\end{eqnarray}
We consider again the case of a narrow voltage pulse applied
at $t=t_0$ and assume a specific shape $\phi(t) = 2n \arctan
[(t-t_0)/\delta t]$. Choosing times $t_1 < t_0 < t_2$ before
and after the application of the pulse at $t_0$, we find that
the currents in the leads `u' and `d' predominantly flow in opposite
directions: for a sharp pulse with $|t_{1,2}-t_0| \gg \delta t$
we can assume that $\phi(t_1) - \phi(t_2) = 2\pi n$ and thus
the correlator ${\cal C}_\mathrm{ud} (t_1,t_2)$ takes the form
\begin{eqnarray}
      &&{\cal C}_\mathrm{ud}(t_1,t_2) =
      \frac{e^2}{\pi^2 (\delta t)^2} T_\mathrm{u} T_\mathrm{d}
      \nonumber\\
      &&\times
      \left[
      \frac{n^2}{(1+z_1^2)(1+z_2^2)}-\frac{\sin^2 \pi n}{(z_1-z_2)^2}
      \right],
\end{eqnarray}
where we have introduced $z_{1,2}=(t_{1,2}-t_0)/\delta t$. For
$|z_{1,2}|\gg 1$ the second (negative) term $\propto 1/(|z_1|
+z_2)^2$ describing the irreducible correlations dominates
over the first (positive) term $\propto 1/(z_1^2\, z_2^2)$
and hence the full current cross-correlator is negative. This
negative sign tells us that, despite application of a positive
voltage pulse with  $n>0$, the currents at times $t_1 < t_0
< t_2$ in leads `u' and `d' flow in opposite directions on
average. Note that this unusual behavior is a specific
feature of time dependent voltage pulses and does not
appear for a constant dc voltage with  $\phi(t) = eVt/\hbar$
--- in this case the current cross-correlator is
always positive.

As a consequence, the time-integrated full particle-number
cross-correlator (per one spin component) may turn out
negative as well and it does so for voltage pulses carrying
less then one electron per spin channel $n<1$,
\begin{equation}
      {\cal K}_\mathrm{ud}^\mathrm{ex} =
      e^2 T_\mathrm{u} T_\mathrm{d} \Bigl[
      n^2 - n -
      \frac{2}{\pi^2}\sin^2(\pi n)\,\ln
      \frac{t_\mathrm{m}}{\delta t}\Bigr].
\end{equation}
Hence, in the weak pumping regime the particles in the
outgoing leads `u' and `d' are preferentially transmitted in
opposite directions. Note that both (negative) terms in the
correlator, the one ($-n$) from short time differences as
well as the contribution ($\propto -\sin^2(n\pi)\ln
(t_\mathrm{m}/\delta t)$ related to the `orthogonality
catastrophe' dominate over the (positive) product term
($n^2$), with the second one becoming increasingly important
at large measuring times. Furthermore, this second term
also drives the particle-number cross-correlator negative
at large non-integer $n>1$ and long measuring times, again
signalling the presence of particle backflow in the device.

The derivation of the Bell inequality relies on
the assumption that the quantities $|x/X|,~|y/Y|$ etc.\ are
bounded by unity. For our setup this implies that the
particle number ratios of the type $|x/X|=|(N_1-N_3)/
(N_1+N_3)|$ are bounded by unity, which is only guaranteed
for particle numbers with equal sign, $N_1\, N_3 >0$;
hence, particles detected in the pair of spin filters
with polarization $\pm{\bf a}$ in the upper arm have to
be transmitted in the same direction. Next, we note that
particles with opposite spin propagate independently and
hence our finding that particles preferentially propagate
in opposite directions of the outgoing leads `u' and `d'
for $n < 1$ also implies that the particle numbers $N_1$
and $N_3$ can be of opposite sign, hence the condition
$x < 1$ is not necessarily satisfied for $n < 1$.
On the contrary, for $n=1$ the full particle-number
cross-correlator (for one spin component) vanishes,
${\cal K}_\mathrm{ud}=0$: in the simplest interpretation
we may conclude that the single transmitted particle
is propagating either through the upper or the lower
lead, thus either $N_\mathrm{u} =0$ and $N_\mathrm{d}=1$
or vice versa and the quantity $|x/X|$ is properly bounded.
The above arguments cannot exclude the relevance of
additional many-particle effects, i.e., the appearance of
additional particle-hole excitations in the system
contributing to the particle count in the various
detectors. A formal proof of the unidirectional
propagation of particles confirming the applicability
of the Bell inequality for the present non-stationary
situation relies on the calculation of the full
counting statistics of particles measured in the
detectors 1 and 3, etc.; such a calculation has not
been done yet.

\subsection{Many integer-flux pulses}\label{sec:mifp}

Above, we have concentrated on the situation where only
a single voltage pulse has been applied. Let us consider
another situation where a sequence of voltage pulses driving
an integer number of electrons is applied to the source lead
`s'.  In contrast to the previous analysis, we study the
total transmitted charge from $t=-\infty$ to $t=\infty$, $\hat
N_i(\infty)= \int_{-\infty}^\infty dt \, \hat I_i(t)$; the excess
part of the irreducible particle-number cross-correlator
takes the form [we remind that the equilibrium part can be
quenched in going to the four-terminal beam splitter of
Fig.\ \ref{fig:fork}(b)]
\begin{equation}
      K^\mathrm{ex}_\mathrm{ud} = -
      \frac{e^2}{\pi^2}T_\mathrm{u} T_\mathrm{d}\!\!
      \int\limits_{-\infty}^{\infty} \!\! dt_1 dt_2\,
      \frac{\sin^2[(\phi(t_1)-\phi(t_2))/2]}{(t_1-t_2)^2}.
      \label{ncordi}
\end{equation}
In our further analysis, we closely follow the technique
developed in Ref.\ \onlinecite{ll_95}. The double
integral in the above expression is logarithmically
divergent at large times $t_1,t_2$, producing the logarithmic
dependence on the measurement time $t_\mathrm{m}$ noted above
for the finite accumulation time. However, for pulses with an
integer number of electrons, this problematic term disappears;
in this case we are allowed to regularize the integral in
(\ref{ncordi}) with the help of
\begin{equation}
      \frac{1}{(t_1-t_2)^2}
      \rightarrow \frac12 \Bigl[ \frac1{(t_1-t_2+i\delta)^2}+
      \frac1{(t_1-t_2-i\delta)^2}\Bigr]
      \label{reg}
\end{equation}
and $\delta \to 0$ a small cutoff. Expressing the factor
$\sin^2(...)$ in Eq.~(\ref{ncordi}) in terms of exponential
functions, we arrive at the form
\begin{eqnarray}
      K_\mathrm{ud}^\mathrm{ex} &=& \frac{e^2
      T_\mathrm{u} T_\mathrm{d}}{(2\pi)^2} \int\!\! dt_1 dt_2 \,
      \biggl[\frac{e^{i\phi(t_1)-i\phi(t_2)}}{(t_1-t_2+i\delta)^2}
      \label{K_ex} \\
      &&\qquad\qquad\qquad\qquad\qquad
      +\frac{e^{i\phi(t_1)-i\phi(t_2)}}{(t_1-t_2-i\delta)^2}
      \biggr].\nonumber
\end{eqnarray}
In order to proceed further, we split the exponential into two
terms, $e^{i\phi(t)} = f_+(t)+f_-(t)$, with $f_+(t)$ and $f_-(t)$
two bounded analytic functions in the upper and lower complex-$t$
plain. Substitution into the above expression and using Cauchy's
formula for the derivative,
\[
   \dot f_\pm(t) = \pm \frac{i}{2\pi}
   \int dt' \frac{f_\pm(t')}{(t-t'\pm i\delta)^2},
\]
allows us to write the particle-number correlator in the form
\begin{equation}
      K_\mathrm{ud}^\mathrm{ex} = - \frac{e^2}{2\pi i}
      T_\mathrm{u} T_\mathrm{d} \int dt\, \bigl[\dot f_+(t)
      f_+^*(t) - \dot f_-(t) f_-^*(t) \bigr].
      \label{ncorf}
\end{equation}
In (\ref{ncorf}) we have made use of the analytical properties of
$f_\pm(t)$; in particular, with the complex conjugate functions
$f^*_+(t)$ and $f^*_-(t)$ bounded and analytic in the lower
and upper half-planes, respectively, we easily find that
$\int dt \, \dot f_+(t) f_-^*(t) = \int dt \, \dot f_-(t)
f_+^*(t) = 0$. In addition, we can also express the average
number of transmitted particles in terms of the functions
$f_\pm(t)$ introduced above,
\begin{eqnarray}
      \langle \hat N_\mathrm{u(d)} \rangle &=&
      \frac{e}{\pi} T_\mathrm{u(d)} \int dt\, \dot \phi(t)
      \label{nave}
      \\
      &=& \frac{e}{\pi i} T_\mathrm{u(d)} \int dt\, e^{-i\phi(t)}
      \frac{d}{dt} e^{i\phi(t)}
      \nonumber\\
      &=& \frac{e}{\pi i} T_\mathrm{u(d)} \int dt\,
      \bigl[ \dot f_+(t) f_+^*(t) + \dot f_-(t) f_-^*(t) \bigr].
      \nonumber
\end{eqnarray}
Rewriting Eqs.~(\ref{ncorf}) and (\ref{nave}) in terms of the
real numbers $n_\pm$,
\begin{equation}
      n_\pm = \pm \frac1{2\pi i} \int dt\, \dot f_\pm(t) f_\pm^*(t),
      \label{npm}
\end{equation}
we obtain the particle-number cross-correlator and the average number
of transmitted particles in the form
\begin{eqnarray}
      K_\mathrm{ud}^\mathrm{ex} &=& - e^2\, T_\mathrm{u}
      T_\mathrm{d}\, (n_+ + n_-),
      \label{ncorn} \\
      \langle N_\mathrm{u(d)}\rangle &=& 2e\,
      T_\mathrm{u(d)}\,(n_+-n_-).
      \label{naven}
\end{eqnarray}
Substituting these expressions into the Bell inequality
Eq.~(\ref{BI3}), we arrive at the result
\begin{equation}
      E_{\scriptscriptstyle\rm BI} = \left|
      \frac{n_+ + n_-}{ 2 (n_+ - n_-)^2 - (n_+ + n_-)}
       \right|\leq \frac1{\sqrt{2}}.
      \label{BI_n}
\end{equation}
The physical meaning of the numbers $n_\pm$ is easily identified
for the specific form of Lorentzian voltage pulses
\begin{equation}
      V(t) = \sum_i n_i\,
      \frac{2\hbar\gamma_i/e}{1+(t-t_i)^2\gamma_i^2},
      \label{vpl}
\end{equation}
where the index $i$ denotes the number of the pulse in the sequence,
$t_i$ is the moment of its appearance, $\gamma_i^{-1}$ is the
pulse width, and $n_i$ the number of spinless electrons carried
by the $i$-th pulse with the sign of $n_i$ defining the sign
of the applied voltage. Such a sequence of pulses produces
the phase
\begin{equation}
      e^{i\phi(t)} = \prod_i
      \Bigl(\frac{t-t_i-i/\gamma_i}{t-t_i+i/\gamma_i}\Bigr)^{n_i},
      \label{ephi}
\end{equation}
from which the decomposition into the terms $f_\pm(t)$ can be
found. The further analysis is straightforward for unidirectional
pulse sequences with all $n_i > 0$, in which case $\exp[i\phi(t)]
= f_+(t)$ and $n_+ = \sum_i n_i$, $n_- = 0$, or all $n_i < 0$
whence $\exp[i\phi(t)] = f_-(t)$ and $n_+ = 0$, $n_- = - \sum_i
n_i$. It then turns out \cite{ll_95,lll_96} that all results
for the irreducible particle-number cross-correlator
(\ref{ncorn}), the average currents (\ref{naven}),
and the Bell inequality (\ref{BI_n}) do neither depend on
the separation $t_{i+1} - t_i$ between the pulses nor
on their widths $\gamma_i^{-1}$. Furthermore, the result
(\ref{BI_n}) for the Bell inequality agrees with the
previous expression (\ref{BI5}) where a single pulse
is carrying $n=n_+$ (or $n_-$) electrons in one go and
we confirm our finding that the violation of the Bell
inequality is restricted to pulses containing only one
pair of electrons with opposite spin. Also, we note that
for the case of well separated pulses we can restrict
the accumulation time over the duration of the
individual pulses, in which case the Bell inequality
is violated for all pulses with $|n_i| = 1$.

Another remark concerns the case of an alternating
voltage signal with no net charge transport and hence
zero accumulated particle numbers $\langle N_{\mathrm{u(d)}}
\rangle = 0$. Equation (\ref{naven}) then tells us
that $n_+ = n_-$ and the Bell inequality (\ref{BI_n})
is formally violated. However, we argue that this
violation is again unphysical and due to the same
improper normalization of the basic quantities
$|x/X|,~|y/Y|$, etc.\ as encountered previously
for the case of small Faraday flux $n < 1$:
concentrating on the expression $x/X
= (N_1-N_3)/(N_1+N_3)$, we note that two pulses
with opposite signs allow for processes where the
charge driven through the two spin detectors satisfies
$N_1 N_3 < 0$ and hence $|x/X| > 1$, in contradiction
with the requirements of the lemma. Note that the manner
of violating the Bell inequality is quite different for the
physical cases involving pulses with a single particle
(see the discussion of single integer-flux
pulses in  Sec.\ \ref{sec:if} with $n=1$, or the
discussion of many integer-flux pulses in Sec.\
\ref{sec:mifp} with $n_+ = 1,~n_-=0$ and $n_+=0,~n_-
= 1$) and for the unphysical situation of an
alternating signal with $n_+ = n_-$ discussed above:
in the first case the small denominator results
from a cancellation between the product term $2n$ and
the negative number correlator $-1$, hence $E_{\rm
\scriptscriptstyle BI} = |1/(2-1)| = 1$, while in the
second case, the product term vanishes and there is no
compensation, although the final result is the same,
$E_{\rm \scriptscriptstyle BI} = |1/-1| = 1$. The
same apparent violation appears at large non-integer
values of $n>1$ and long measuring times, where the
term $\propto \sin^2(\pi n)\ln(t_\mathrm{m}/\delta t)$
becomes dominant, cf.\ (\ref{BI4}).

\section{Conclusion}\label{sec:conc}

The application of voltage pulses to a mesoscopic fork
allows to generate spin-entangled pairs of electrons
through post-selection; the presence of these entangled pairs
can be observed in a Bell inequality measurement based on
particle-number cross-correlators. A number of items have
to be observed in producing these entangled objects:
{\it i)} Equilibrium fluctuations competing with the
pulse signal have to be eliminated. This can be achieved
with the help of a four-channel beam splitter as sketched
in Fig.\ \ref{fig:fork}(b) where the channel mixing is
tuned such that the transmission $T_\mathrm{ud}$ between
the upper and lower channel is blocked.
{\it ii)} Pulses $V(t)$ with integer Faraday flux $\Phi
= -c \int dt  V(t) = n \Phi_0$ injecting an integer number
of particles shall be used. Otherwise, the `fractional
injection' of a particle induces a long-time perturbation
in the system producing a logarithmically divergent
contribution to the excess number correlator. The flux
$\Phi = (n+\delta n)\Phi_0$, $n = $ an integer, associated
with the voltage pulse has to be precise within the limit
$(\delta n)^2 \ll 1/\ln(t_m/\delta t)$, with $t_m$ the
measurement time of the pulse and $\delta t$ the pulse width.
{\it iii)} The Bell inequality is violated for
pulses injecting a single pair of electrons with opposite
spin, i.e., pulses with one Faraday flux and hence $n=1$.
The maximal violation of the inequality points to the
full entanglement of the pair --- the question what
type of pulses produce only partially entangled states
(as quantified in terms of concurrence or negativity
of the partially transposed density matrix \cite{peres_96})
has not been addressed here.
{\it iv)} Although weak pumping with pulses carrying less
than one Faraday flux, i.e., $n <1$, formally violate
the Bell inequality (note the proviso {\it ii)}, however),
we associate this spurious violation with an improper
normalization of the particle-number ratios $(N_i-N_j)
/(N_i+N_j)$ entering the Bell inequality.
{\it v)} The same argument also applies to the case of
pumping with an alternating signal --- we find the Bell
inequality always violated when the average injected
current vanishes (i.e., when the number of carriers
transmitted in the forward and backward directions are
equal). Again, the origin of this spurious violation is
located in the improper normalization of the particle-number
ratios $(N_i-N_j)/(N_i+N_j)$ for this situation.

The above points suggest the following physical interpretation:
An integer-flux pulse with Faraday flux $n\Phi_0$ extracts
exactly $n$ electron pairs from the reservoir which then
are tested in the Bell measurement setup. For $n=1$ we find
the Bell inequality maximally violated, implying that the
electrons within the pair are maximally entangled and not
entangled with the remaining electrons in the
Fermi sea. On the other hand, the application of a
fractional-flux pulse with non-integer $n$ produces
a superposition of states with different number of
excess electron pairs in the fork. The electrons
injected into the fork then remain entangled with
those in the Fermi sea and their analysis in the
Bell measurement setup makes no sense.

In our analysis of the spurious violations of Bell
inequalities for weak pumping and for alternating drives we
have identified the presence of reverse particle flow as
the problematic element.  In the weak pumping limit this
conclusion has been conjectured from the appearance of
negative values in the current cross-correlator, implying
negative values of the particle-number correlator for $n
< 1$. Although we believe that these are strong arguments
supporting our interpretation, we are not aware of a formal
analysis of the backflow appearing in this type of systems.
The question to be addressed then is: Given a bias signal
driving particles through the device in the forward direction,
what are the circumstances and what is the probability to
find particles moving in the opposite direction (backflow)?
A related problem has been addressed by Levitov
\cite{levitov_01} (see also Ref.\ \onlinecite{makhlin_01})
who has derived the full counting statistics for the charge
transport across a quantum point contact in the weak $ac$
pumping regime and has identified parameters producing a
strictly unidirectional flow. The corresponding analysis
for our system remains to be done.

A similar scheme for producing spin-entangled pairs of
electrons has been discussed in Ref.~\onlinecite{lebedev_05},
where a constant voltage $V$ has been applied to the source
lead `s'. In this case, the source reservoir injects a regular
sequence of spin-singlet pairs of electrons separated by the
voltage time $\tau_V = h/eV$; the
Bell inequality then is violated at short times only. The main
novelty of the present proposal is the generation of
well separated spin-entangled electron pairs in response to
distinguished voltage pulses, thus avoiding the short time
correlation measurement at time scales of order
$\tau_V$.

Our mesoscopic fork device produces entangled pairs of
electrons with a probability of 50 \%, i.e., half of the
single-flux pulses will produce a useful pair with one
spin propagating in the upper and the other in the lower
channel. The competing events with both particles moving
in one channel produce no useful outcome. This is similar
to the finding of Beenakker {\it et al.} \cite{beenakker_05}
who derive a concurrence corresponding to the production
of one entangled pair per two pumping cycles. In how
far this represents an upper limit in the performance
of this type of devices or what type of entanglement generators
are able to reach (at least ideally) 100 \% efficiency is
an interesting problem.

We thank D. A. Ivanov and L. S. Levitov for discussions and
acknowledge the financial support from the Swiss National
Foundation (through the program MaNEP and the CTS-ETHZ), the
Forschungszentrum J\"ulich within the framework of the Landau
Program, the Russian Science Support Foundation, and the program
'Quantum Macrophysics' of the RAS.


\begin{thebibliography}{99}
\bibitem{note} The setup in Ref.\ \onlinecite{lebedev_05} involves
   a short-time measurement (on a time scale $\tau \sim h/eV$)
   with well separated detectors $x+y \ll \tau v_F$; under these
   conditions the contribution of the equilibrium part of the 
   correlator is small compared to the excess part.
\end{thebibliography}
\end{document}